\magnification=1200
\baselineskip=13pt
\overfullrule=0pt
\tolerance=100000

\font\tenbifull=cmmib10 \skewchar\tenbifull='177
\font\tenbimed=cmmib7   \skewchar\tenbimed='177
\font\tenbismall=cmmib5  \skewchar\tenbismall='177
\textfont9=\tenbifull
\scriptfont9=\tenbimed
\scriptscriptfont9=\tenbismall

\mathchardef\alpha="710B
\mathchardef\beta="710C
\mathchardef\gamma="710D
\mathchardef\delta="710E
\mathchardef\epsilon="710F
\mathchardef\zeta="7110
\mathchardef\eta="7111
\mathchardef\theta="7112
\mathchardef\iota="7113
\mathchardef\kappa="7114
\mathchardef\lambda="7115
\mathchardef\mu="7116
\mathchardef\nu="7117
\mathchardef\micron="716F
\mathchardef\xi="7118
\mathchardef\pi="7119
\mathchardef\rho="711A
\mathchardef\sigma="711B
\mathchardef\tau="711C
\mathchardef\upsilon="711D
\mathchardef\phi="711E
\mathchardef\chi="711F
\mathchardef\psi="7120
\mathchardef\omega="7121
\mathchardef\varepsilon="7122
\mathchardef\vartheta="7123
\mathchardef\varphi="7124
\mathchardef\varrho="7125
\mathchardef\varsigma="7126
\mathchardef\varpi="7127

 at 8truept

{\hfill \hbox{\vbox{\settabs 1\columns
\+ solv-int/9601001\cr
}}}
\centerline{}
\bigskip
\bigskip
\bigskip
\baselineskip=18pt

\centerline{\bf Hamiltonian Structures for the Generalized Dispersionless
KdV Hierarchy}
\vfill

\centerline{J. C. Brunelli\footnote{*}{Work supported by CNPq. E-mail address: jbrunelli@if.usp.br}}
\medskip
\medskip
\centerline{Departamento de F\'\i sica Matem\'atica}
\centerline{Instituto de F\'\i sica da Universidade de S\~ao Paulo}
\centerline{C.P. 66318, CEP 05389-970}
\centerline{S\~ao Paulo, SP - BRAZIL}
\vfill

\centerline{\bf {Abstract}}

\medskip
\medskip

We study from a Hamiltonian point of view the generalized
dispersionless KdV hierarchy of equations. From the so called
dispersionless Lax representation of these equations we obtain three
compatible Hamiltonian structures. The second and third Hamiltonian
structures are calculated directly from the r-matrix approach. Since the third
structure is not related recursively with the first two ones 
the generalized dispersionless KdV hierarchy can be characterized as
a truly tri-Hamiltonian system.
\vfill
\eject

\bigskip
\leftline{\bf 1. Introduction}
\medskip

The representation of nonlinear integrable equations by a system of
linear equations is due to Lax and have been studied extensively in
the past [1-3]. An interesting class of nonlinear equations are
the so called dispersionless Lax equations which are the quasi-classical limit
of ordinary Lax equations. This quasi-classical limit corresponds to
the solutions which slowly depend on the variables $x,t$. Let us take
the KdV equation
$$
4u_t=u_{xxx}+6uu_x\eqno(1.1)
$$
and after dropping out the dispersive term (or make the substitution
${\partial\ \over\partial t}\to\epsilon{\partial\ \over\partial
t},{\partial\ \over\partial x}\to\epsilon{\partial\ \over\partial x}$
and $\epsilon\to0$) we end up with the equation
$$
u_t={3\over2}u u_x\eqno(1.2)
$$
This is the Riemann equation (also called inviscid Burgers', Hopf
etc.). Solutions of (1.2) can be written through the implicit form [4]
$$
u=f(x-ut)\eqno(1.3)
$$
and this dependence gives rise to the breaking of the wave shape
leading to a transition from conservative to dissipative behaviour
[5]. As it is well known  the balancing between the dispersive and nonlinear
terms in (1.1) is responsible for the soliton solutions and the
integrability of (1.1). What is interesting is that the evolution
equation (1.2), at least before the breaking of its wave solutions, is a
integrable Hamiltonian system much like the KdV system (1.1). In fact
this behaviour was observed for the dispersionless equations
describing an incompressible nonviscous fluid with a free surface in
the approximation of long waves, the Benney's equation [6]. A
Hamiltonian structure for this equation was obtained and studied by
Kupershmidt and Manin [7], Manin [8], Lebedev and Manin [9], and Lebedev
[10]. As point out by Olver and Nutku [11] equation (1.2) has an infinite
sequence of zero order (the highest order derivative which a
$F(u,u_x,\dots)$ depends) conserved charges
$$
H_{1}=\int dx\,u\,,\quad
H_{3}={1\over4}\int dx\,u^2\,,\quad
H_{5}={1\over8}\int dx\,u^3\,,\dots\eqno(1.4)
$$
and has three first-order Hamiltonian structures given by
$$
\eqalign{
{\cal D}_1=&2\partial\cr
{\cal D}_2=&u\partial +\partial u\cr
{\cal D}_3=&u^2\partial+\partial u^2\cr
\
}\eqno(1.5)
$$
These three Hamiltonian operators are compatible in the Magri's sense
[12,13] making (1.2) a tri-Hamiltonian system, i.e., equation (1.2) can be
written in three Hamiltonian forms
$$
u_t={\cal D}_1{\delta H_5\over\delta u}=
{\cal D}_2{\delta H_3\over\delta u}=
{3\over4}{\cal D}_3{\delta H_1\over\delta u}\eqno(1.6)
$$
However, it is important to point out that since
$$
{\cal D}_3\not={\cal D}_2{\cal D}_1^{-1}{\cal D}_2\eqno(1.7)
$$
the Hamiltonian operators are not trivially related. Therefore, the
Hamiltonian structures (1.5) make the Riemann equation a truly
tri-Hamiltonian system. This is to be contrasted with the
Kupershmidt's equations for the dispersive water wave equations [14]. In
that situation the three Hamiltonian structures are related by a
unique recursion operator.

It follows, for instance from Magri's theorem [12,13], that the
Hamiltonians (1.4) are in involution with respect to any of the three
Poisson brackets
$$
\{H_n,H_m\}_i=\int dx\,{\delta H_n\over\delta u}{\cal D}_i{\delta
H_m\over\delta u}=0\,,\quad i=1,2,3,\dots\eqno(1.8)
$$
Whence, (1.2) is an integrable Hamiltonian system.

We can associate with (1.2) a Lax representation using the Chen et al
approach [15]. The Lax pair is
$$
{\partial L\over\partial t}=[L,B]\eqno(1.9)
$$
where $L$ is the recursion operator obtained from (1.5)
$$
L={\cal D}_2{\cal D}_1^{-1}=u+\partial u\partial^{-1}\eqno(1.10)
$$
and $B$ is the Frech\'et derivative obtained through a linearization
of the Riemann equation (1.2)
$$
v_t={3\over2}{d\ \over d\epsilon}(u+\epsilon v)(u+\epsilon
v)_x\Big|_{\epsilon=0}={3\over2}\partial uv=Bv
\eqno(1.11)
$$
Also, there is another Lax representation (1.9) with
$$
\eqalign{
L=&\partial+u\cr
B=&{3\over4}u^2\cr
}\eqno(1.12)
$$
Although we have  Lax representations for the Riemann
equation we do not know how to use them for an inverse scattering
problem or use the pseudo-differential operator algebra to obtain
its Hamiltonian structures through the Gelfand-Dickey approach [3]. 
However, Lebedev [10] has notice that for the case of Benney's
equation an alternative Lax representation is possible. This Lax
representation is called dispersionless Lax equation and was also
considered by Krichever [16] in his studies about topological minimal
models. In this paper we will use this dispersionless Lax
representation and the algebraic setup behind it to derive the
Hamiltonian structures of the Riemann equation and the hierarchy of
equations which contains it as the first nontrivial equation. This
paper is organized as follows. In section 2 we introduce the
generalized dispersionless KdV hierarchy of equations and we show that
the Riemann equation belongs to it. In section 3 we obtain the first
Hamiltonian structure. The second and third Hamiltonian structures for the
generalized dispersionless KdV hierarchy are obtained in section 4 by
the r-matrix method. We present our conclusions in section 5.
\bigskip
\leftline{\bf 2. Dispersionless KdV Hierarchy}
\medskip

The dispersionless KdV equation (1.2) can be obtained directly,
bypassing the dispersionless limit of (1.1).
Let $E_n$ be the polynomial of degree $n$ in $p$ [16]
$$
E_n=p^n+u_{-1}p^{n-1}+u_0p^{n-2}+\cdots+u_{n-2}=\sum_{i=0}^n
u_{n-i-2}p^i
\eqno(2.1)
$$
where $u_{-2}=1$ and the polynomial coefficients $u_i$ are functions
of the variable $x$ and various time variables $t_k$
($k=1,2,3,\dots$). We denote $A_+$ and $A_-$ the parts of the Laurent
polynomial $A$ containing nonnegative and negative powers of $p$
respectively. The generalized dispersionless KdV hierarchy is given by
the Lax equation 
$$
{\partial E_n\over\partial t_k}=\{(E_n^{k/n})_+,E_n\}=\{E_n,(E_n^{k/n})_-\}\eqno(2.2)
$$
where the bracket is defined [9,10] to be
$$
\{A,B\}=\partial_p A\,\partial_x B-\partial_p B\,\partial_x A\eqno(2.3)
$$
and $E_n^{k/n}$ is the $k$th power of the Laurent polynomial $E_n^{1/n}$
satisfying $(E_n^{1/n})^n=E_n$.

First, let us denote the degree of a Laurent polynomial of the form
$A=\sum_i a_ip^i$ as
$$
\hbox{deg}\,A\equiv \hbox{maximum $i$ for which $a_i\not=0$}\eqno(2.4)
$$
and observe that the bracket reduces the degree of a polynomial by one
unit. Therefore,
$$
\hbox{deg}\,\{A,B\}=\hbox{deg}\,A+\hbox{deg}\,B-1\eqno(2.5)
$$
From (2.2) we have
$$
\hbox{deg}\,{\partial E_n\over\partial t_k}=
\hbox{deg}\,\{E_n,(E_n^{k/n})_-\}\le n-1-1=n-2\eqno(2.6)
$$
and when we compare with (2.1) we conclude that
$$
{\partial u_{-1}\over\partial t_k}=0\eqno(2.7)
$$
for any $k$.

Now, for general Laurent polynomials of the form
$$
A=\sum_{i=-\infty}^{+\infty} a_i(x)p^i\eqno(2.8)
$$
we define the residue as the coefficient of the $p^{-1}$ term
$$
\hbox{Res}A=a_{-1}\eqno(2.9)
$$
and the Adler trace [17] as
$$
\hbox{Tr}\,A=\int dx\,\hbox{Res}A\eqno(2.10)
$$
For general Laurent polynomials $A=\sum_{i=-\infty}^{+\infty} a_ip^i$
and $B=\sum_{i=-\infty}^{+\infty} b_ip^i$ it is straightforward to show
that
$$
\hbox{Res}\{A,B\}=\sum_{i=-\infty}^{+\infty} i(a_ib_{-i})'\eqno(2.11)
$$
which implies
$$
\hbox{Tr}\,\{A,B\}=0\eqno(2.12)
$$
Also, it follows easily from (2.12) the useful property 
$$
\hbox{Tr}\,(A\{B,C\})=\hbox{Tr}\,(B\{C,A\})\eqno(2.13)
$$

Let us note from (2.2) that we can write
$$
{\partial E_n^{m/n}\over\partial t_k}=\{(E_n^{k/n})_+,E_n^{m/n}\}
\eqno(2.14)
$$
for an arbitrary integer $m$. Taking the trace of (2.14) and after
using (2.12) we obtain
$$
{\partial\ \over\partial t_k}\hbox{Tr}\,(E_n^{m/n})=0\eqno(2.15)
$$
Thus, we define the conserved charges as
$$
H_m={n\over m}\hbox{Tr}\,(E_n^{m/n})\,,\quad m=1,2,3\dots\eqno(2.16)
$$

Now let us show that the flows given by (2.2) commute. Let be
$$
\eqalign{
{\partial^2 E_n\over\partial t_\ell\partial t_k}=&{\partial\ \over\partial
t_\ell}\{(E_n^{k/n})_+,E_n\}\cr
=&\{({\partial E_n^{k/n}/\partial
t_\ell})_+,E_n\}+\{(E_n^{k/n})_+,\{(E_n^{\ell/n})_+,E_n\}\}\cr
=&\{\{(E_n^{\ell/n})_+,E_n^{k/n}\}_+,E_n\}+\{(E_n^{k/n})_+,\{(E_n^{\ell/n})_+,E_n\}\}\cr
}\eqno(2.17)
$$
where we have used (2.14). Now using the Jacobi identity
$$
\eqalign{
{\partial^2 E_n\over\partial t_\ell\partial t_k}
=&\{\{(E_n^{\ell/n})_+,E_n^{k/n}\}_+,E_n\}+
\{\{(E_n^{k/n})_+,(E_n^{\ell/n})_+\},E_n\}
+\{(E_n^{\ell/n})_+,\{(E_n^{k/n})_+,E_n\}\}\cr
=&\{\{(E_n^{\ell/n})_+,(E_n^{k/n})_-\}_+,E_n\}+
\{(E_n^{\ell/n})_+,\{(E_n^{k/n})_+,E_n\}\}\cr
=&\{\{E_n^{\ell/n}- (E_n^{\ell/n})_-,(E_n^{k/n})_-\}_+,E_n\}+
\{(E_n^{\ell/n})_+,\{(E_n^{k/n})_+,E_n\}\}\cr
=&\{\{E_n^{\ell/n},E_n^{k/n}-(E_n^{k/n})_+\}_+,E_n\}+
\{(E_n^{\ell/n})_+,\{(E_n^{k/n})_+,E_n\}\}\cr
=&\{\{(E_n^{k/n})_+,E_n^{\ell/n}\}_+,E_n\}+\{(E_n^{\ell/n})_+,\{(E_n^{k/n})_+,E_n\}\}\cr
=&{\partial^2 E_n\over\partial t_k\partial t_\ell}\cr
}
\eqno(2.18)
$$
Therefore, the hierarchy of equations (2.2) has an infinite number of
conserved laws (2.16) and an infinite number of commuting flows and
can be formally be considered integrable.

Let us illustrate these results for (2.1) with $n=2$ and $u_{-1}=0$. We
obtain for $E_2\equiv E$ and $u_0\equiv u$
$$
\eqalign{
E=&p^2+u\cr
E^{1/2}=&p+{1\over2}up^{-1}-{1\over8}u^2p^{-3}+{1\over16}u^3p^{-5}
-{5\over128}u^4p^{-7}+\dots\cr
E^{3/2}=&p^3+{3\over2}up+{3\over8}u^2p^{-1}+\dots\cr
E^{5/2}=&p^5+{5\over2}up^3+{15\over8}u^2p+{5\over16}u^3p^{-1}+\dots\cr
E^{7/2}=&p^7+{7\over2}up^5+{35\over8}u^2p^3+{35\over16}u^3p+{35\over128}u^4p^{-1}+\dots\cr
{}\vdots&\cr
}\eqno(2.19)
$$
From
$$
{\partial E\over\partial t_k}=\{(E^{k/2})_+,E\}\eqno(2.20)
$$
we get the hierarchy of equations
$$
\eqalign{
{\partial u\over \partial t_1}=&u_x\cr
{\partial u\over \partial t_3}=&{3\over2}uu_x\cr
{\partial u\over \partial t_5}=&{15\over 8}u^2u_x\cr
{\partial u\over \partial t_7}=&{35\over 16}u^3u_x\cr
{}\vdots&\cr
}\eqno(2.21)
$$
The conserved charges from (2.16) are
$$
H_m={2\over m}\hbox{Tr}\,(E^{m/2})\eqno(2.22)
$$
and results in
$$
\eqalign{
H_1=&\int dx\, u\cr
H_3=&{1\over4}\int dx\, u^2\cr
H_5=&{1\over8}\int dx\, u^3\cr
H_7=&{5\over64}\int dx\, u^4\cr
{}\vdots&\cr
}\eqno(2.23)
$$
Thus, from (2.21) we get the Riemann equation as the first
nonlinear equation in the hierarchy. Also, the charges (2.23) are
exactly the Hamiltonians (1.4). Then we will call (2.20) and (2.2)
the dispersionless KdV (Riemann) hierarchy and the generalized
dispersionless KdV (Riemann) hierarchy, respectively.

\bigskip
\leftline{\bf 3. First Hamiltonian Structure}

Now we will derive in a systematic way the first Hamiltonian structure
associated with the generalized dispersionless KdV hierarchy of equations
(2.2). Particularly, as an example we will obtain the first
Hamiltonian structure in (1.5) for the Riemann equation. Here we
will follow the Drinfeld and Sokolov approach [18] as in [19] for the
usual KdV hierarchy.

Let us observe that
$$
(E_n^{k/n})_-=E_n^{k/n}-(E_n^{k/n})_+=\sum_{i=1}^{n-1}\hbox{Res}\left(E_n^{k/n}p^{i-1}\right)p^{-i}+{\cal
O}(p^{-n})\eqno(3.1)
$$
and even though $(E_n^{k/n})_-$ has an infinite number of terms the
only ones that will contribute to give dynamical equations are the
ones up to degree $p^{-n+1}$. Since
$$
p^i=\int dx\,{\delta E_n\over\delta u_{n-i-2}}\eqno(3.2)
$$
we rewrite (3.1) as
$$
\eqalign{
(E_n^{k/n})_-=&\sum_{i=1}^{n-1}\hbox{Res}\left(E_n^{k/n}\int dx\,{\delta
E_n\over\delta u_{n-i-1}}\right)p^{-i}+{\cal O}(p^{-n})\cr
{}=&\sum_{i=1}^{n-1}{\delta H_{n+k}\over\delta u_{n-i-1}}p^{-i}+{\cal O}(p^{-n})\cr
}\eqno(3.3)
$$
where $H_{n+k}$ is given by (2.16).

Using (3.3) in (2.2) we get
$$
{\partial E_n\over\partial t_k}=\{E_n,\sum_{i=1}^{n-1}{\delta H_{n+k}\over\delta u_{n-i-1}}p^{-i}\}\eqno(3.4)
$$
since the terms of order ${\cal O}(p^{-n})$ in (3.3) do not
contribute. Let us introduce the dual to $E_n$
$$
Q=\sum_{i=1}^n q_{n-i-1}p^{-i}\eqno(3.5)
$$
where the $q$'s are assumed to be independent of the $u$'s. This
yields a linear functional
$$
\hbox{Tr}\,E_nQ=\int dx\,\sum_{i=0}^{n-1} u_{n-i-2}q_{n-i-2}\eqno(3.6)
$$
Also, let us note that
$$
\sum_{i=1}^{n-1}{\delta H_{n+k}\over\delta u_{n-i-1}}p^{-i}=
\sum_{i=1}^{n-1}\int dy\,{\delta H_{n+k}\over\delta u_{n-i-1}(y)}V_{i}(x,y)\eqno(3.7)
$$
where
$$
V_{i}(x,y)\equiv \delta(x-y)p^{-i}\eqno(3.8)
$$
and which gives
$$
\hbox{Tr}\,E_nV_{i}(x,y)=u_{n-i-1}(y)\eqno(3.9)
$$
We have thus from (3.4) using (3.9) and (2.13)
$$
{\partial\ \over\partial t_k}\hbox{Tr}\,E_nQ=\sum_{i=1}^{n-1}\int
dy\,\hbox{Tr}\,E_n\{V_{i}(x,y),Q(x)\}{\delta H_{n+k}\over\delta
u_{n-i-1}(y)}\eqno(3.10)
$$
If we want  to write this equation in Hamiltonian form as
$$
\eqalign{
{\partial\ \over\partial
t_k}\hbox{Tr}\,E_nQ=&\{\hbox{Tr}\,E_nQ,H_{n+k}\}_1\cr
=&\sum_{i=1}^{n-1}\int dy\,\{\hbox{Tr}\,E_nQ,\hbox{Tr}\,E_nV_{i}\}_1{\delta
H_{n+k}\over\delta u_{n-i-1}(y)}\cr
}\eqno(3.11)
$$
(where we have used (3.9)) we get after comparing it with (3.10)
$$
\{\hbox{Tr}\,E_nQ,\hbox{Tr}\,E_nV_{i}\}_1=\hbox{Tr}\,E_n\{V_i,Q\}\eqno(3.12)
$$
In this way, the dispersionless Lax equation (2.2) can be written in
Hamiltonian form with respect to the first Poisson bracket
$$
\{\hbox{Tr}\,E_nQ,\hbox{Tr}\,E_nV\}_1=\hbox{Tr}\,E_n\{V,Q\}\eqno(3.13)
$$
for any dual $Q$ and $V$ relative to $E_n$. As an example let be
$$
E_2\equiv E=p^2+u_{-1}p+u_0\eqno(3.14)
$$
with the duals
$$
\eqalign{
V=&v_{-1}\,p^{-2}+v_{0}\,p^{-1}\cr
Q=&q_{-1}\,p^{-2}+q_{0}\,p^{-1}\cr
}\eqno(3.15)
$$
We get that
$$
\hbox{Tr}\,E\{V,Q\}=2\int dx\, q_0v_0'\eqno(3.16)
$$
and
$$
\eqalign{
\hbox{Tr}\,EQ=&\int dx\,(u_{-1}q_{-1}+u_0q_0)\cr
\hbox{Tr}\,EV=&\int dx\,(u_{-1}v_{-1}+u_0v_0)\cr
}\eqno(3.17)
$$
Now, (3.13) yields
$$
\eqalign{
\{u_{-1}(x),u_{-1}(y)\}_1=&0\cr
\{u_{-1}(x),u_{0}(y)\}_1=&0\cr
\{u_{0}(x),u_{0}(y)\}_1=&2\partial\delta(x-y)\cr
}\eqno(3.18)
$$
and constraining (3.14) to $E=p^2+u$, where $u\equiv u_0$, we must
set $u_{-1}=0$. We finally get from (3.18)
$$
\{u(x),u(y)\}_1=2\partial\delta(x-y)={\cal D}_1\delta(x-y)\eqno(3.19)
$$
which is the first Hamiltonian structure for the Riemann equation
(1.2). 

We now could try to write the dispersionless Lax equation in
other Hamiltonian forms. However, in the next
section we will use the algebraic structure behind the dispersionless
hierarchy and apply the r-matrix
formalism to obtain the other Hamiltonian structures.
\vfill\eject
\bigskip
\leftline{\bf 4. Second and Third Hamiltonian Structures: r-Matrix Approach}
\medskip

It is well known by now that the so called first Hamiltonian structure
of integrable models is the sympletic structure of Kostant-Kirillov
[20] on the orbits of the coadjoint representation of Lie groups
[17,21]. For dispersionless equations given by the Lax representation
(2.2) the corresponding Lie algebra is given by the associative
algebra of Laurent polynomials endowed with the bracket (2.3). The
Hamiltonian structure (3.13) can also be obtained from this
result. For Benney's equation this interpretation was already given by
Lebedev [10] and Lebedev and Manin [9].

Semenov-Tian-Shansky [22] has shown that the multi-Hamiltonian nature
of integrable equations could be explained in terms of the so called
r-matrix. Let $\hbox{g}$ be an abstract associative algebra with a
non-degenerate trace form $\hbox{Tr}\colon\hbox{g}\to\hbox{\bf R}$. In this way
we can identify $\hbox{g}$ with its dual $\hbox{g}^*$ by
$$
\langle A\vert B\rangle=\hbox{Tr}\,AB\eqno(4.1)
$$
Also, we can use the natural Lie algebra structure obtained by
$[A,B]=AB-BA$ on $\hbox{g}$. The linear mapping
$R\colon\hbox{g}\to\hbox{g}$ is a classical r-matrix on $\hbox{g}$ whenever
the modified bracket
$$
[A,B]_R=[RA,B]+[A,RB]\eqno(4.2)
$$
satisfies the Jacobi identity [22]. This gives us a second Lie algebra
structure on $\hbox{g}$. The bracket (4.2) satisfies the Jacobi
identity if the modified Yang-Baxter equation holds. The important
result for us is that the new Lie product endows $\hbox{g}=\hbox{g}^*$ with
new Poisson structures. The first one is
$$
\{\hbox{Tr}\,EQ,\hbox{Tr}\,EV\}_2={1\over 2}\hbox{Tr}\,E
\Bigl([Q,R(EV)]+[R(QE),V]\Bigr)\eqno(4.3)
$$
where $Q$ and $V$ are duals to $E$. This expression was given by
Semenov-Tian-Shansky (formula (22) in [22]) and it is the analog of
the second structure of Gelfand-Dickey [3]. In references [23] and
[24] a third Poisson structure was introduced
$$
\{\hbox{Tr}\,EQ,\hbox{Tr}\,EV\}_3={1\over 2}\hbox{Tr}\,E
\Bigl([Q,R(EVE)]+[R(EQE),V]\Bigr)\eqno(4.4)
$$
and it was shown that the Kostant-Kirillov structure (which gives
(3.13) in our problem), (4.3) and (4.4) form a compatible
tri-Hamiltonian system, i.e., the three structures are compatible in
Magri's sense.

For Lie algebras that can be written in the form
$$
\hbox{g}=\hbox{g}_+\oplus\hbox{g}_-\eqno(4.5)
$$
the r-matrix on $\hbox{g}$ is given by
$$
R=P_+-P_-\eqno(4.6)
$$
where $P_\pm\hbox{g}=\hbox{g}_\pm$ are the projections onto the
subalgebras. For our particular case of dispersionless equations it is
clear that
$$
\eqalign{
\hbox{g}_+=&\left\{ A_+=\sum_{i=0}^{\infty}a_i(x)p^i\right\}\cr
\hbox{g}_-=&\left\{ A_-=\sum_{i=1}^{\infty}a_{-i}(x)p^{-i}\right\}\cr
}\eqno(4.7)
$$
with trace given by (2.6) and bracket given by (2.3). So, the Poisson brackets
(4.3) and (4.4) assume the form
$$
\displaylines{
\quad\{\hbox{Tr}\,E_nQ,\hbox{Tr}\,E_nV\}_2=\hfill\cr
\hfill={1\over 2}\hbox{Tr}\,E_n
\Bigl(\{Q,(E_nV)_+\}-\{Q,(E_nV)_-\}+\{(E_nQ)_+,V\}-\{(E_nQ)_-,V\}\Bigr)\hfill(4.8a)\cr
\quad\{\hbox{Tr}\,E_nQ,\hbox{Tr}\,E_nV\}_3=\hfill
\cr
\hfill={1\over 2}\hbox{Tr}\,E_n
\Bigl(\{Q,(E_n^2V)_+\}-\{Q,(E_n^2V)_-\}+\{(E_n^2Q)_+,V\}-\{(E_n^2Q)_-,V\}\Bigr)\hfill(4.8b)\cr
}
$$

Let us again use the Riemann equation as an example. Using (3.14) and
(3.15) we get, after  straightforward algebra, the following Poisson
brackets from (4.8)
$$
\eqalign{
\{u_{-1}(x),u_{-1}(y)\}_2=&-2\partial\delta(x-y)\cr
\{u_{-1}(x),u_{0}(y)\}_2=&-\partial u_{-1}\delta(x-y)\cr
\{u_{0}(x),u_{0}(y)\}_2=&\left(u_{0}\partial+ \partial u_{0}-
u_{-1}\partial u_{-1}\right)\delta(x-y)\cr
}\eqno(4.9a)
$$
and
$$
\eqalign{
\{u_{-1}(x),u_{-1}(y)\}_3=&-2\left(u_{0}\partial+\partial u_{0}\right) \delta(x-y)\cr
\{u_{-1}(x),u_{0}(y)\}_3=&-\left(2\partial u_{0}u_{-1}+
u^2_{-1}\partial u_{-1}\right)\delta(x-y)\cr
\{u_{0}(x),u_{0}(y)\}_3=&\left(u^2_{0}\partial+\partial
u^2_{0}-u_{-1}\partial u_{0}u_{-1}-u_{0}u_{-1}\partial
u_{-1}\right)\delta(x-y)\cr
}\eqno(4.9b)
$$
From the first Poisson bracket (3.18) we see that $u_{-1}$ decouples
from $u_0$. However, in the second and third brackets (4.9) $u_{-1}$
is coupled to itself and to $u_0$. From $\{u_{-1}(x),u_{-1}(y)\}$ in
(4.9) it follows that $u_{-1}=0$ corresponds to a second class
constraint and we have to use the Dirac reduction (see [24] for
example). We then obtain
$$
\eqalignno{
\{u(x),u(y)\}_2=&(u\partial+\partial u)\delta(x-y)={\cal
D}_2\delta(x-y)&(4.10a)\cr
\{u(x),u(y)\}_3=&(u^2\partial+\partial u^2)\delta(x-y)={\cal
D}_3\delta(x-y)&(4.10b)\cr
}
$$ 
where we have set $u\equiv u_0$. These are exactly the Hamiltonian
structures in (1.5).

Finally, the $k$th flow in the generalized dispersionless KdV hierarchy
(2.2) can be written in Hamiltonian form as
$$
{\partial {\bf u}\over\partial t_k}
={\cal D}_1{\delta H_{k+n}\over\delta {\bf u}}
={\cal D}_2{\delta H_{k}\over\delta {\bf u}}
={k(k-2)\over (k-1)^2}{\cal D}_3{\delta H_{k-n}\over\delta {\bf u}}
\eqno(4.11)
$$
where $k>1$ and ${\bf u}=(u_{-1},u_0,\dots,u_{n-2})$. The Hamiltonians $H_n$ are given by
(2.16) and the Hamiltonian structures ${\cal D}_1$, ${\cal D}_2$ and
${\cal D}_3$ can be obtained from the Poisson brackets  (3.13), (4.8a)
and (4.8b) respectively. For $n=2$ we obtain from (4.11) the
dispersionless KdV hierarchy of equations (2.21).
                                                                                                                                                                                                                                       
\bigskip
\leftline{\bf 5. Conclusions}
\medskip

We have studied the Hamiltonian structures of the generalized
dispersionless KdV hierarchy of equations. We have obtained the second
and third Hamiltonian structures directly from the r-matrix approach
following the results of Semenov-Tian-Shansky [22], Lin and Parmentier
[23] and Oevel and Ragnisco [24]. We have illustrated our main results
through the Riemann equation (1.2). However, as was discussed by Olver
and Nutku the Riemann equation has an additional Hamiltonian
structure. Namely, the third-order Hamiltonian
$$
{\cal E}=\partial{1\over u_x}\partial{1\over u_x}\partial\eqno(5.1)
$$
allows us to write the dispersionless KdV hierarchy of equations in
Hamiltonian form. This Hamiltonian structure is only compatible with the
${\cal D}_1$ structure. So, the Riemann equation and the higher order
equations (2.21) are quadri-Hamiltonian systems. Consequently, this
property should be valid for the whole generalized KdV hierarchy of
equations. We did not succeeded in deriving a Poisson bracket, yielding
(5.1) for the Riemann equation, from the r-matrix approach. This is an
interesting problem and is under investigation.

With the results obtained in this paper we can study the higher
Hamiltonian structures of other interesting dispersionless
systems. One of them is the so called Benney [6] system of
equations. This system has a dispersionless Lax representation much
like the generalized dispersionless KdV hierarchy of equations. For
instance, the classical dispersionless long wave equation
$$
\eqalign{
u_t+uu_x+h_x=&0\cr
h_t+(uh)_x=&0\cr
}\eqno(5.2)
$$
can be derived from the Benney's system. It is not difficult to check
that it has a simple dispersionless nonstandard Lax representation
$$
{\partial E\over\partial t}=\{E,(E^2)_{\ge1}\}\eqno(5.3)
$$
where
$$
E=p+{1\over2}u+{1\over4}hp^{-1}
\eqno(5.4)
$$
and the bracket is given by (2.3). Here $(E^2)_{\ge1}$ stands for the
purely nonnegative (without $p^0$ terms) part of the Laurent
polynomial obtained from $E^2$. For dispersive systems the
nonstandard Lax representation was introduced by Kupershmidt in [14]
and the generalization of the Gelfand-Dikii brackets was performed in
[25]. The derivation of the Poisson brackets for equations with
nonstandard dispersionless Lax representation is an interesting and
relevant problem and is also under investigation.
\bigskip
\leftline{\bf Acknowledgements}
\medskip
 
I would like to thank Ashok Das for discussions as well as for helpful
comments.
\vfill\eject

\leftline{\bf References}
\bigskip

\item{1.} L. D. Faddeev and L. A. Takhtajan, ``Hamiltonian Methods in 
the Theory of Solitons'' (Springer, Berlin, 1987).

\item{2.} A. Das, ``Integrable Models'' (World Scientific, Singapore, 
1989).

\item{3.} L. A. Dickey, ``Soliton Equations and Hamiltonian Systems'' (World
Scientific, Singapore, 1991).

\item{4.} G. B. Whitham, ``Linear and Nonlinear Waves'' (Wiley, New
York, 1974).

\item{5.} P. Lax, Am. Math. Mo. {\bf 79}, 227 (1972); J. Cavalcante and H. P. McKean, Physica {\bf 4D}, 253 (1982).

\item{6.} D. J. Benney, Studies Appl. Math. {\bf 52}, 45 (1973).

\item{7.} B. A. Kupershmidt and Yu. I. Manin, Funct. Anal. Appl. {\bf
11}, 188 (1977); {\bf 12}, 20 (1978).

\item{8.} Yu. I. Manin, J. Sov. Math. {\bf 11}, 1 (1979).

\item{9.} D. R. Lebedev and Yu. I. Manin, Phys. Lett. {\bf 74A}, 154
(1979).

\item{10.}D. R. Lebedev, Lett. Math. Phys. {\bf 3}, 481 (1979). 

\item{11.} P. J. Olver and Y. Nutku, J. Math. Phys. {\bf 29}, 1610 (1988).

\item{12.} F. Magri, J. Math. Phys. {\bf 19}, 1156 (1978).

\item{13.} P. J. Olver , ``Applications of Lie Groups to Differential
Equations'', Graduate Texts in Mathematics, Vol. 107 (Springer, New York,
1986).

\item{14.} B.A. Kupershmidt, Comm. Math. Phys. {\bf 99}, 51 (1985).

\item{15.} H. H. Chen, Y. C.  Lee and C. S. Liu, Phys. Scripta {\bf
20}, 490 (1979); J. C. Brunelli and A. Das, Mod. Phys. Lett. {\bf A10},
931 (1995).

\item{16.} I. Krichever, Commun. Math. Phys. {\bf 143}, 415 (1992).

\item{17.} M. Adler, Invent. Math. {\bf 50}, 219 (1979).

\item{18.} V. G. Drinfeld and V. V. Sokolov, J. Sov. Math. {\bf 30},
1975 (1985).

\item{19.} A. Das and W.-J. Huang, J. Math. Phys. {\bf 33}, 2487
(1992).

\item{20.} A. A. Kirillov, ``Elements of the Theory of
Representations'' (Springer, Berlin, 1976); B. Kostant, Lect. Notes in
Math. {\bf 170}, 87 (1970); J. M. Souriau, ``Structure des Systems
Dynamiques (Dunod, Paris, 1970).

\item {21.} B. Kostant, London Math. Soc. Lect. Notes, Ser. {\bf 34},
287 (1979); A. G. Reyman, M. A. Semenov-Tian-Shansky and I. B. Frenkel,
Sov. Math. Dokl. {\bf 20}, 811 (1979); A. G. Reyman and
M. A. Semenov-Tian-Shansky, Invent. Math. {\bf 54}, 81 (1979); {\bf
63}, 423 (1981); W. W. Symes, Invent. Math. {\bf 59}, 13 (1980);
D. R. Lebedev and Yu. I. Manin, Funct. Anal. Appl. {\bf 13}, 268 (1980).
 
\item{22.} M. A. Semenov-Tian-Shansky, Funct. Anal. Appl. {\bf 17}, 259 (1983).

\item{23.} L.-C. Li and S. Parmentier,  Comm. Math. Phys. {\bf 125},
545 (1989).

\item{24.} W. Oevel and O. Ragnisco, Physica {\bf 161A}, 181 (1990).

\item{25.} J. C. Brunelli, A. Das and W.-J. Huang,
Mod. Phys. Lett. {\bf A9}, 2147 (1994). 
\end